\documentclass[%
 aip,
 amsmath,amssymb,
 reprint,%
]{revtex4-1}
\usepackage{graphicx}% Include figure files
\usepackage{dcolumn}% Align table columns on decimal point
\usepackage{bm}% bold math
\usepackage[utf8]{inputenc}
\usepackage[T1]{fontenc}
\usepackage{mathptmx}
\usepackage{float}
\usepackage{upgreek}
\usepackage{color}
\usepackage{booktabs}
\usepackage{multirow}
\usepackage{siunitx}
\begin{document}

\preprint{AIP/123-QED}

\title{Sublattice mixing in Cs$_2$AgInCl$_6$ for enhanced optical properties from first-principles}
% Force line breaks with \\

\author{Manish Kumar}
 \email{manish.kumar@physics.iitd.ac.in}
% \altaffiliation[Also at ]{Physics Department, XYZ University.}%Lines break automatically or can be forced with \\
\author{Manjari Jain}
\author{Arunima Singh}
\author{Saswata Bhattacharya}%
 \email{saswata@physics.iitd.ac.in}
\affiliation{ 
Department of Physics, Indian Institute of Technology Delhi, New Delhi 110016, India%\\This line break forced with \textbackslash\textbackslash
}%

%\author{C. Author}
% \homepage{http://www.Second.institution.edu/~Charlie.Author.}
%\affiliation{%
%Second institution and/or address%\\This line break forced% with \\
%}%

%\date{\today}% It is always \today, today,
             %  but any date may be explicitly specified

\begin{abstract}
Lead-free double perovskite materials (viz. Cs$_2$AgInCl$_6$) are being explored as stable and non-toxic alternatives of lead halide perovskites. In order to expand the optical response of Cs$_2$AgInCl$_6$ in visible region, we report here the stability, electronic structure and optical properties of Cs$_2$AgInCl$_6$ by sublattice mixing of various elements. Here, we have employed %high-throughput screening using
a hierarchical first-principles based approach starting from density functional theory (DFT) with appropriate exchange-correlation functionals to beyond DFT methods under the framework of many body perturbation theory (viz. G$_0$W$_0$@HSE06). We have started with 32 primary set of combinations of metals M(I), M(II), M(III) and halogen X at Ag/In and Cl site, respectively, where concentration of each set is varied to build a database of nearly 140 combinations. The most suitable mixed sublattices are identified to engineer the band gap of Cs$_2$AgInCl$_6$ to have its application in optoelectronic devices under visible light.
\end{abstract}

\maketitle

\section{Introduction}
Lead halide perovskites APbX$_3$ (A = ${\textrm{CH}_3\textrm{NH}_3\mathstrut}^+$, ${\textrm{HC}(\textrm{NH}_2)_2\mathstrut}^+$, ${\textrm{Cs}\mathstrut}^+$, and X = ${\textrm{Cl}\mathstrut}^-$, ${\textrm{Br}\mathstrut}^-$, ${\textrm{I}\mathstrut}^-$) have created a huge sensation in the field of optoelectronics, particularly in photovoltaics owing to their suitable optical band gap, long carrier diffusion length, high carrier mobility and low manufacturing cost
~\cite{doi:10.1021/ja809598r,Lee643,doi:10.1002/adma.201305172,doi:10.1021/jz500858a,pooja-prb, doi:10.1021/acs.chemrev.6b00136}. Moreover, the band gap is tunable with high defect tolerance~\cite{doi:10.1021/acsenergylett.6b00196,tan2018dipolar}. These materials find applications in various optoelectronic devices, namely, solar cells~\cite{Lee643,burschka2013sequential,doi:10.1021/jz5011187}, light emitting diodes~\cite{doi:10.1021/acs.nanolett.5b00235,doi:10.1021/acs.jpclett.5b02011}, lasers~\cite{doi:10.1021/jz5005285,xing2014low} and photodetectors~\cite{doi:10.1021/acsnano.6b08194,doi:10.1002/adma.201803422}. 
In spite of their great potential in vast number of applications, there are two major challenges: (i) instability against exposure to humidity, heat or light and (ii) toxicity of Pb. To tackle these issues, many works have been endeavored to find the alternative stable and environmentally sustainable metal halide perovskites with fascinating optoelectronic properties akin to lead halide perovskites~\cite{C4TA05033A,doi:10.1021/acs.chemmater.5b01989,C6CC06475B,doi:10.1021/acsenergylett.7b01197,C7TA08992A}.

One of the approaches for removing Pb-toxicity is to replace ${\textrm{Pb}\mathstrut}^{2+}$ with some other divalent metal. However, this replacement results in either indirect or large band gap materials with degraded optoelectronic properties~\cite{C5TC04172D,doi:10.1021/acs.jpcc.5b06436,PhysRevB.94.180105}. Substitution of group 14 divalent cations, viz. ${\textrm{Sn}\mathstrut}^{2+}$ and ${\textrm{Ge}\mathstrut}^{2+}$ have also been synthesized by researchers, but these are not stable at ambient conditions due to the easy oxidation to tetravalent ${\textrm{Sn}\mathstrut}^{4+}$ and ${\textrm{Ge}\mathstrut}^{4+}$, respectively~\cite{doi:10.1021/jacs.5b13470,doi:10.1021/acsenergylett.7b00191}. Another promising approach is to 
substitute a monovalent M(I) and a trivalent M(III) metal alternatively in place of two divalent Pb, that forms the double perovskite A$_2$M(I)M(III)X$_6$. Many high-throughput calculations have been performed on double perovskites for a variety of potential applications~\cite{C9TA01456J,doi:10.1021/acs.chemmater.9b00116,D0TC02231D}. Recently, lead-free metal halide double perovskites have been synthesized, which are stable and environmentally benign, viz. Cs$_2$AgBiX$_6$ (X = ${\textrm{Cl}\mathstrut}^-$, ${\textrm{Br}\mathstrut}^-$, ${\textrm{I}\mathstrut}^-$) and Cs$_2$AgInCl$_6$~\cite{doi:10.1021/jacs.5b13294,doi:10.1021/acs.chemmater.5b04231,doi:10.1021/acs.jpclett.6b00376,doi:10.1021/acs.jpclett.6b01041,doi:10.1021/acs.jpclett.6b02682,C7TA04690A,doi:10.1021/jacs.8b07983}. Cs$_2$AgBiX$_6$ perovskites possess indirect band gap, which results in weaker absorption and high non-radiative recombination loss~\cite{doi:10.1021/jacs.5b13294,doi:10.1021/acs.chemmater.5b04231,doi:10.1021/acs.jpclett.6b00376}. In contrast, Cs$_2$AgInCl$_6$ has direct band gap and long carrier lifetimes. However, its wide band gap (3.3 eV) does not show optical response in visible region~\cite{doi:10.1021/acs.jpclett.6b02682,C7TA04690A}. Alloying with suitable elements could be the best solution to reduce its band gap and expand the spectral response in visible light region. In recent studies, Cs$_2$AgInCl$_6$ has been doped to tune its optical properties~\cite{C8CC01982G,doi:10.1021/jacs.8b07983,doi:10.1021/acs.chemmater.9b00410,doi:10.1021/acs.chemmater.9b02973,doi:10.1002/anie.202002721}.

In this Letter, we have done the sublattice mixing by partial substitution of several metals M(I), M(II), M(III) and halogen X at Ag/In and Cl site, respectively, to reduce the band gap of Cs$_2$AgInCl$_6$, thereby, enhancing its optical properties. The charge neutrality condition has been maintained by forming substitutional defects. We have performed %high-throughput screening by performing
hierarchical calculations using first-principles based approaches viz. density functional theory (DFT) with semi-local exchange-correlation (xc) functional (PBE~\cite{PhysRevLett.77.3865}), hybrid DFT with HSE06~\cite{doi:10.1063/1.1564060,doi:10.1063/1.2404663} and single shot GW~\cite{PhysRev.139.A796,PhysRevLett.55.1418} (G$_0$W$_0$) under the many body perturbation theory (MBPT). Firstly, the structural stability analysis has been done by examining the Goldschmidt tolerance factor and octahedral factor. Since, structural stability is not the sufficient condition to confirm the formation of perovskites, the decomposition energy~\cite{doi:10.1002/anie.201705113} has been calculated, which reflects the thermodynamic stability of the materials. We have taken the difference between total energy of the configurations and their components (binary/ternary, in which they can decompose), which is opposite to what has been considered in Ref.~\cite{doi:10.1002/anie.201705113} Therefore, the configurations that have negative decomposition energy are stable. Further, to get better insights, we have investigated the reduction in band gap via atom projected partial density of states. Finally, by calculating the frequency dependent complex dielectric function, we have determined the optical properties of the materials that can be applied in the field of optoelectronics.

\section{Methods}
The DFT calculations have been performed using the Vienna \textit{ab initio} simulation package (VASP)~\cite{KRESSE199615}. The ion-electron interactions in all the elemental constituents are described using projector-augmented wave (PAW) potentials~\cite{PhysRevB.50.17953}. All the structures are optimized  %(i.e., the atomic positions are relaxed)
using generalized gradient approximation (PBE xc functional) until the forces are smaller than 0.001 eV/\AA. Here, the PBE xc functional is used because the HSE06 xc functional is extremely slow for relaxing the structure. In case of double perovskite Cs$_2$AgInCl$_6$, the lattice constant is overestimated by 1.56\% using PBE xc functional (and by 1.22\% using HSE06 xc functional) in comparison to experimental value obtained by Volonakis \textit{et al}~\cite{doi:10.1021/acs.jpclett.6b02682}. Whereas, the PBEsol xc functional underestimates the lattice constant by 1\%. The electronic self-consistency loop convergence is set to 0.01 meV, and the kinetic energy cutoff used is 500 eV for plane wave basis set convergence. A $k$-mesh of $4\times4\times4$ is used for Brillouin zone integration, which is generated using Monkhorst-Pack~\cite{PhysRevB.13.5188} scheme. Advanced hybrid xc functional HSE06 is used for the better estimation of band gap as well as thermodynamic stability. Further, we have checked the role of van der Waals (vdW) forces and configurational entropy, while analyzing the stability of compounds. The latter has lesser effect on the stability. The consideration of vibrational energy contributes to second decimal place of the decomposition energy. It may change the number by very small amount, but neither changes the stability nor the hierarchy of stability of the compounds. On the other hand, the van der Waals forces (two-body Tkatchenko-Scheffler~\cite{PhysRevLett.102.073005}) contribute to first decimal place of the decomposition energy. Most of the compounds' stability has not been affected. However in very few cases, it has minutely changed the stability of the compounds, that have decomposition energy value close to zero. For determination of the optical properties, single shot GW (G$_0$W$_0$) calculations have been performed on top of the orbitals obtained from HSE06 xc functional [G$_0$W$_0$@HSE06]. The polarizability calculations are performed on a grid of 50 frequency points. The number of bands is set to four times the number of occupied bands (for band gap convergence see Table S1 in Supporting Information (SI)). %The spin-orbit coupling (SOC) is not taken into account due to its negligible affect on the electronic structure of Ag/In-based halide double perovskite~\cite{C7MH00239D,doi:10.1021/acs.chemmater.9b00116}.
Moreover, the negligible effect of spin-orbit coupling (SOC) have been discussed.

\section{Results and Discussion}
\subsection{Stability of defected systems}
\subsubsection{Structural stability}
\begin{figure}
	\centering
	\includegraphics[width=0.35\textwidth]{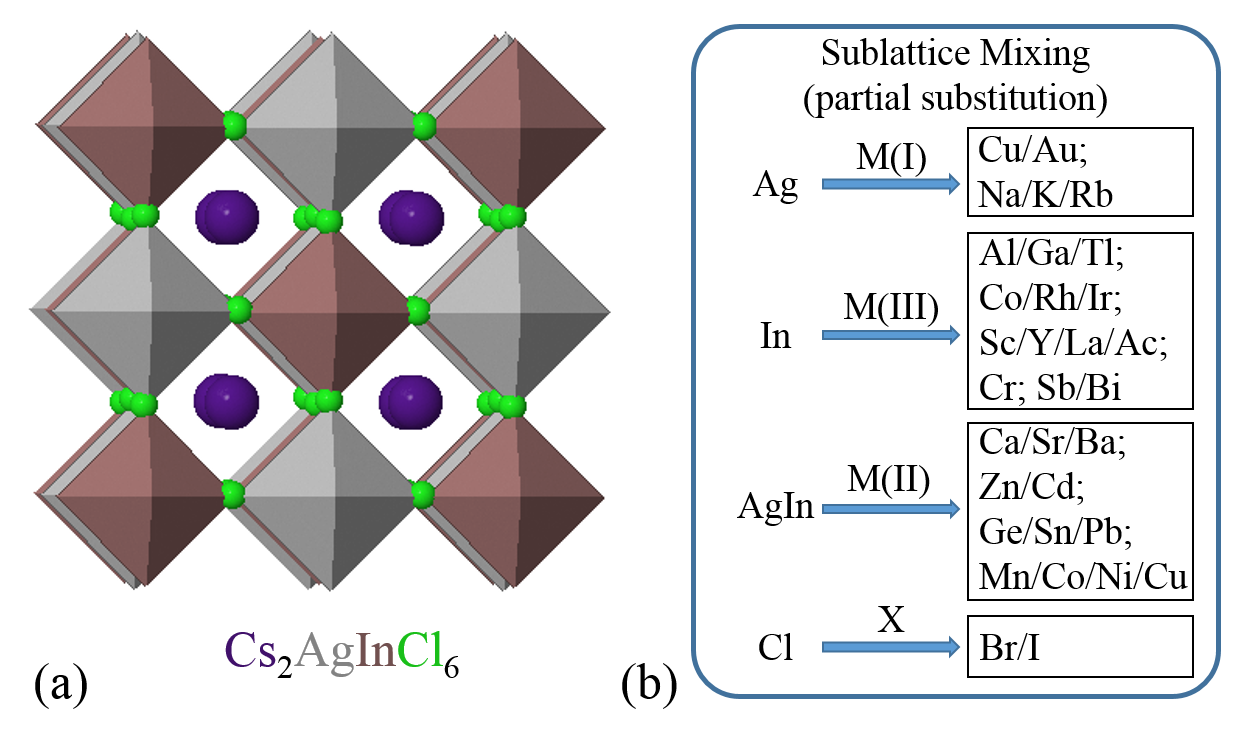}
	\caption{(a) Structure of Cs$_2$AgInCl$_6$, and (b) Partial substitution with metals M(I), M(II), M(III) and with halogen X at Ag/In and Cl site, respectively.}
	\label{fig_dp}
\end{figure}
The double perovskite Cs$_2$AgInCl$_6$ has a cubic structure with space group \textit{Fm$\bar{3}$m}. The corresponding sublattice is composed of alternate octahedra of InCl$_6$ and AgCl$_6$ as shown in Fig.~\ref{fig_dp}(a). On partial substitution of different elements as shown in Fig.~\ref{fig_dp}(b) (metals and/or halogens), the distortion is negligible (see Section II in SI). Here we have started with 32 primary set of combinations of metals M(I), M(II), M(III) and halogen X at Ag/In and Cl site, respectively, where concentration of each set is varied to build a database of nearly 140 combinations. However, note that here, we have presented the results of 25\% substitution for metals and 4\% substitution for halogen atoms. This is because we have seen and thoroughly checked that, with the increase in concentration of the external element, if the band gap is increased (or decreased), the same trend is followed with further increase in concentration. Two such test cases are shown in Fig. S1. We have also reported this, to be the case in our previous experimental finding~\cite{doi:10.1021/acs.jpclett.9b02168}. Moreover, some combinations beyond 25\% substitution are not considered in the following cases: (i) toxic elements (viz. Tl(III), Cd(II), Pb(II)), (ii) elements that lead to instability on 25\% substitution (viz. Co(II), Cu(II), Ni(II)), and (iii) elements that result in larger (larger than pristine Cs$_2$AgInCl$_6$) indirect band gap on 25\% substitution (viz. Ac(III), Ba(II), Ge(II) and Sn(II)). The structural stability of all the configurations has been determined by calculating two geometrical parameters, viz. the Goldschmidt tolerance factor ($t$) and the octahedral factor ($\mu$). For single perovskite ABX$_3$, $t = (r_\textrm{A} + r_\textrm{X})/\sqrt{2}(r_\textrm{B} + r_\textrm{X})$ and $\mu = r_\textrm{B}/r_\textrm{X}$, where $r_\textrm{A}$, $r_\textrm{B}$, and $r_\textrm{X}$ are the ionic radii of cation A, B, and anion X, respectively. In case of double perovskites, $r_\textrm{B}$ is the average of the ionic radii at B sites. For stable perovskites, the ranges of $t$ and $\mu$ are: $0.8\leq t\leq1.0$ and $\mu>0.41$~\cite{C5SC04845A}. The Shannon ionic radii~\cite{doi:10.1107/S0567739476001551} have been considered to evaluate $t$ and $\mu$. For the configurations we have considered, $t$ lies between 0.85 and 0.91, and $\mu$ has the value between 0.50 and 0.59 (see Table S2 in SI). Therefore, these probable structures are stable.

\subsubsection{Thermodynamic stability}
\begin{figure}
	\centering
	\includegraphics[width=0.48\textwidth]{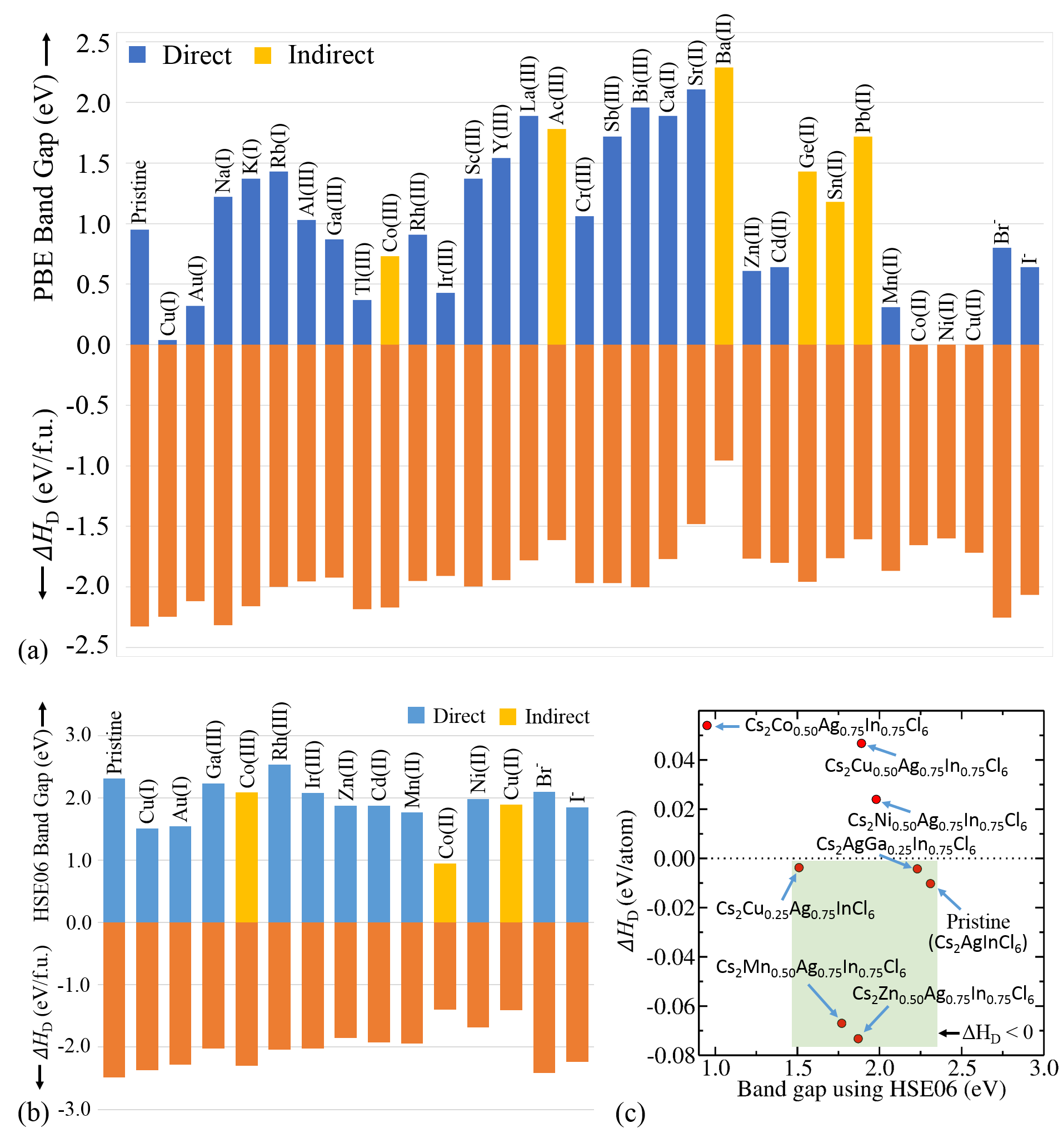}
	\caption{Decomposition energy ($\Delta H_\textrm{D}$) for decomposition of pristine and other configurations into binary compounds, and band gap using the xc functionals (a) PBE and (b) HSE06. (c) Decomposition energy ($\Delta H_\textrm{D}$) for decomposition into ternary compounds using HSE06 xc functional.}
	\label{fig_stable}
\end{figure}
In order to determine the thermodynamic stability, we have computed the decomposition energy ($\Delta H_\textrm{D}$) using PBE and HSE06 xc functionals. We have substituted the external elements in Cs$_8$Ag$_4$In$_4$Cl$_{24}$ supercell framework to model a solid solution.\footnote{As most of the sites in double perovskite are equivalent under same symmetry point, the energy difference in case of different sites (i.e., non-equivalent cells with same stoichiometry) is very small. Thus, even if we take thermodynamic average considering alloying, the results will not get changed.} In order to model the defected system, we have used an iterative procedure as shown in Ref.\cite{pooja-prb,doi:10.1021/acs.jpcc.9b11160}. The $\Delta H_\textrm{D}$ for the decomposition of Cs$_8$Ag$_4$In$_4$Cl$_{24}$ into binary compounds is calculated as follows:
\begin{equation}\begin{split}
\Delta H_\textrm{D}(\textrm{Cs}_8\textrm{Ag}_4\textrm{In}_4\textrm{Cl}_{24}) &= \textrm{E}(\textrm{Cs}_8\textrm{Ag}_4\textrm{In}_4\textrm{Cl}_{24}) - 8\textrm{E}(\textrm{CsCl})
\\ &\quad- 4\textrm{E}(\textrm{AgCl}) - 4\textrm{E}(\textrm{InCl}_3)\textrm{,}
\end{split}\end{equation} 
where $\textrm{E}(\textrm{Cs}_8\textrm{Ag}_4\textrm{In}_4\textrm{Cl}_{24})$, $\textrm{E}(\textrm{CsCl})$, $\textrm{E}(\textrm{AgCl})$, and $\textrm{E}(\textrm{InCl}_3)$ are the DFT energies of the respective compounds. The configurations having negative value of the $\Delta H_\textrm{D}$ are stable. The entropy of mixing is not considered here %, since its contribution is negligible (small fraction of k$_\textrm{B}$T, i.e., few meV) as compared to the decomposition energy (which is 10's or 100's of meV). Therefore, the incorporation of entropy mixing
as it will not change the overall trend i.e., the relative stability will remain same~\cite{doi:10.1021/jacs.9b12440,RevModPhys.86.253,ZHAO20181662}.
%Moreover, the synthesis of halide perovskites is often performed near room temperature, where the 0 K enthalpy of solid-state compounds is a good approximation to predict the thermodynamic stability, hence, the entropy of mixing can be neglected
Fig.~\ref{fig_stable}(a) and~\ref{fig_stable}(b) show the decomposition energy for decomposition of Cs$_2$AgInCl$_6$ and other mixed sublattices into binary compounds using PBE and HSE06 xc functionals, respectively (decomposition pathways are shown in SI). Only those elements, which lead to decrement in band gap using PBE xc functional, are further considered with HSE06 xc functional. %(refer to shaded region in Fig 2(a)).

The quaternary compounds can be decomposed into ternary compounds. Therefore, we have also considered those pathways for the materials [see decomposition pathways which are more probable (as per the smaller value of decomposition energy) in SI], which have the favorable band gap. The decomposition energy for decomposition of Cs$_2$AgInCl$_6$ and other mixed sublattices into ternary compounds is shown in Fig.~\ref{fig_stable}(c). For decomposition of Cs$_8$Ag$_4$In$_4$Cl$_{24}$ into ternary compounds, the $\Delta H_\textrm{D}$ is determined as follows:
\begin{equation}\begin{split}
\Delta H_\textrm{D}(\textrm{Cs}_8\textrm{Ag}_4\textrm{In}_4\textrm{Cl}_{24}) &= \textrm{E}(\textrm{Cs}_8\textrm{Ag}_4\textrm{In}_4\textrm{Cl}_{24}) - 2\textrm{E}(\textrm{CsAgCl}_2)\\&\quad- 2\textrm{E}(\textrm{Cs}_3\textrm{In}_2\textrm{Cl}_9) - 2\textrm{E}(\textrm{AgCl})\textrm{.}
\end{split}\end{equation}
The $\Delta H_\textrm{D}$ has the value of $-2.48$ eV/f.u. and $-0.10$ eV/f.u. for the decomposition of Cs$_8$Ag$_4$In$_4$Cl$_{24}$ into binary and ternary compounds, respectively. These negative values confirm that the perovskite Cs$_8$Ag$_4$In$_4$Cl$_{24}$ is stable.  We have found that all the selected elements for sublattice mixing are stable with respect to the decomposition into binary compounds (see Table S2 in SI). However, for ternary decomposition pathway, Co(II), Ni(II), and Cu(II) are not stable (see Fig.~\ref{fig_stable}(c), where shaded region indicates the stable compounds, i.e., $\Delta H_\textrm{D}<0$). This may be attributed to the smaller size of these cations that are unable to accommodate two octahedra with Cl$_6$, and the lowest octahedral factor of Ni(II) and Cu(II) (see Table S2 in SI). Also, Cu(I) and Ga(III) are less stable than pristine (see Fig.~\ref{fig_stable}(c)). Moreover, we have noticed that Cu(I) is not stable at all (as positive value of $\Delta H_\textrm{D}$ = 0.32 eV/f.u. (see Table~\ref{tbl1})) when it has fully replaced the Ag, i.e., for 100\% substitution, which is in agreement with previous studies~\cite{doi:10.1002/anie.201705113}. It is only stable for 25\% substitution. Therefore, it is concluded that, if the difference between sizes of the substitutional cations/anion and pristine's cations/anion is large, then that configuration would become unstable on increment in concentration.

\begin{table}
\centering
  \caption{Decomposition energy (for the decomposition into ternary compounds) of Cs$_2$Cu$_x$Ag$_{1-x}$InCl$_6$}
	\label{tbl1}
  \begin{tabular}{lSS}
    \toprule
    \multirow{2}{*}{Compounds} &
      \multicolumn{2}{c}{\quad$\Delta H_\textrm{D}$ (eV/f.u.)} \\
      {} & {\; PBE} & {\; HSE06} \\
      \midrule
    Cs$_2$AgInCl$_6$ & -0.098 &  -0.103 \\
    Cs$_2$Cu$_{0.25}$Ag$_{0.75}$InCl$_6$ & -0.070 & -0.039  \\
    Cs$_2$Cu$_{0.50}$Ag$_{0.50}$InCl$_6$ & 0.058 & 0.122 \\
    Cs$_2$Cu$_{0.75}$Ag$_{0.25}$InCl$_6$ & 0.190 & 0.354 \\
    Cs$_2$CuInCl$_6$ & 0.322 & 0.472 \\
    \bottomrule
  \end{tabular}
\end{table}
\subsection{Electronic structure analysis}
\begin{figure}
	\centering
	\includegraphics[width=0.4\textwidth]{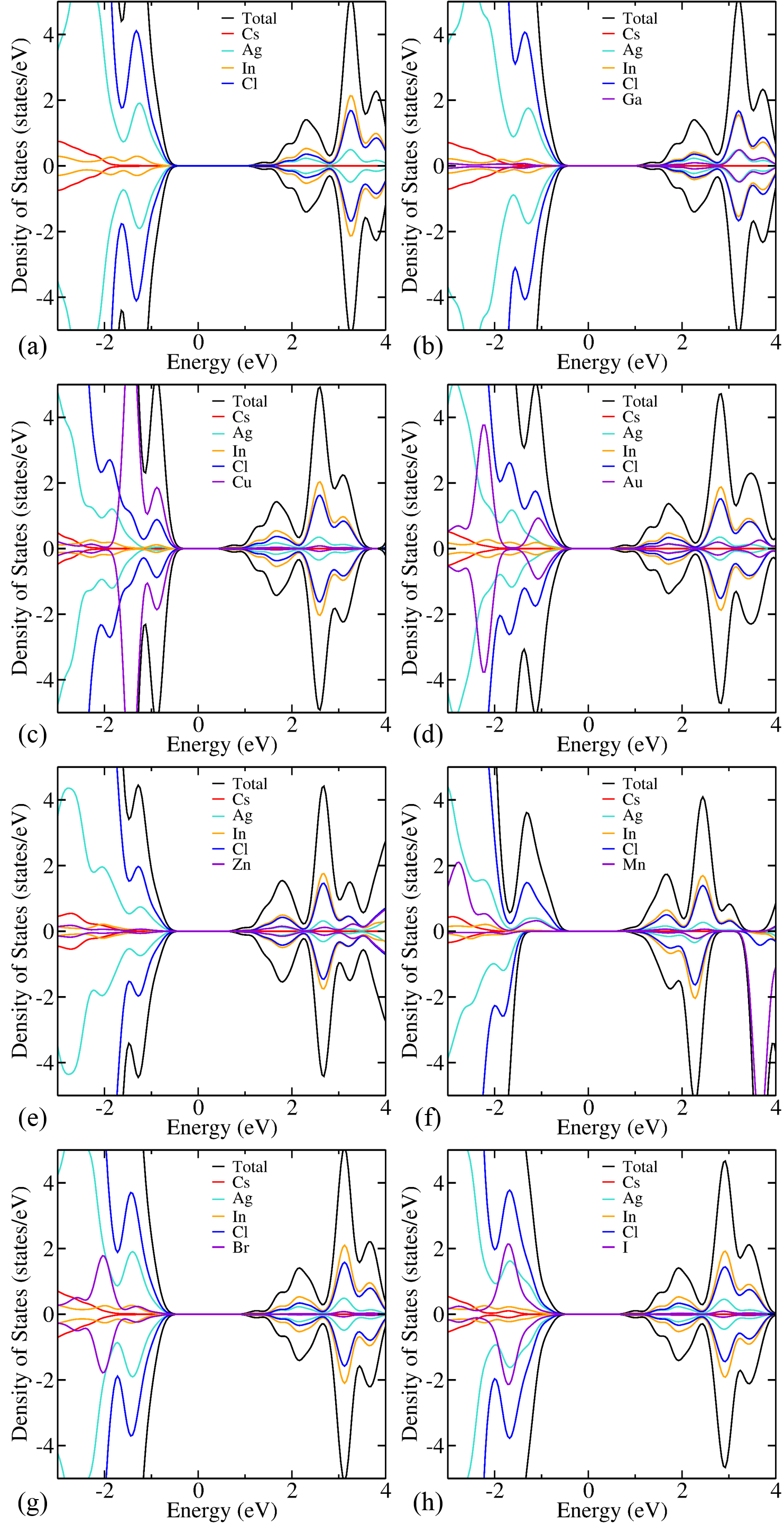}
	\caption{Atom projected partial density of states (pDOS) using HSE06 xc functional of (a) pristine Cs$_2$AgInCl$_6$, (b) Cs$_2$AgGa$_{0.25}$In$_{0.75}$Cl$_6$, (c) Cs$_2$Cu$_{0.25}$Ag$_{0.75}$InCl$_6$, (d) Cs$_2$Au$_{0.25}$Ag$_{0.75}$InCl$_6$, (e) Cs$_2$Zn$_{0.50}$Ag$_{0.75}$In$_{0.75}$Cl$_6$, (f) Cs$_2$Mn$_{0.50}$Ag$_{0.75}$In$_{0.75}$Cl$_6$, (g) Cs$_2$AgInBr$_{0.04}$Cl$_{5.96}$, and (h) Cs$_2$AgInI$_{0.04}$Cl$_{5.96}$.}
	\label{fig_dos}
\end{figure}
A screening of various atoms for sublattice mixing has been done by calculating the band gap first using generalized gradient approximation (PBE) and, subsequently, with inclusion of SOC. The respective band gaps as obtained for pristine Cs$_2$AgInCl$_6$ are 0.95 eV and 0.93 eV, implying insignificant SOC effect on its electronic properties. % This corroborates with the fact that it does not contain any heavy element like Pb or Bi, where some significant effect of SOC is expected in the electronic properties of such materials~\cite{pooja-prb}. In addition, we have also checked the SOC effect in Cs$_2$Au$_{0.25}$Ag$_{0.75}$InCl$_6$. Only the VBM level is lifted by 0.1 eV (see Fig. S3). The nature of the band gap still remains direct at $\Gamma$-point. We have not observed any splitting at band edges, i.e., at conduction band minimum (CBm) or valence band maximum (VBM). The bands remain degenerate at the VBM and CBm.
Also, as per existing literature, SOC has negligible effect in Ag/In~\cite{C7MH00239D,doi:10.1021/acs.chemmater.9b00116} and Au~\cite{PhysRevApplied.13.014005,doi:10.1021/acs.chemmater.6b01348} based double perovskite (see Fig. S3). Therefore, we have ignored the effect of SOC in our further calculations. However, the band gap is highly underestimated by PBE xc functional due to the well-known self-interaction error. Therefore, we have further performed the calculations using hybrid xc functional HSE06 for those mixed sublattices, where in comparison to pristine, the band gap was reduced (see Fig.~\ref{fig_stable}(a),~\ref{fig_stable}(b), and Table S2 in SI). The calculated value of band gap for Cs$_2$AgInCl$_6$ is 2.31 eV using default exact Fock exchange of 25\%, which is in good agreement with previously reported theoretical value, but still underestimated in comparison to the experimental value (3.3 eV)~\cite{doi:10.1021/acs.jpclett.6b02682}. We have also validated that the band gap becomes 3.19 eV on increasing the exact Fock exchange parameter to 40\%. Despite the proximity of this value to that of experiments, it can be drastically changed for the systems having defects (substitution of different elements), and determining it accurately is not possible without the experimental inputs. In view of this, we have used the default 25\% exact Fock exchange parameter for our study assuming this will give atleast the correct trends. In case of Cu(I) and Au(I) substitutional alloying at Ag site, the band gap is reduced by $\sim$ 0.8 eV, having a value of 1.51 eV and 1.54 eV, respectively. On the other hand, in case of substitution of M(III) at In site, it does not have much effect on reduction in band gap. Only Co(III) and Ir(III) substitutional alloying are able to reduce the band gap from 2.31 eV to 2.08 eV, whereas the rest are either increasing it or have no effect on the band gap. In case of M(II) substitutional alloying, one at Ag and other one at In site, only Zn(II) and Mn(II) are able to reduce the band gap effectively, having a band gap value of 1.87 eV and 1.77 eV, respectively. %The Mn(II) doped Cs$_2$AgInCl$_6$ has also been realized experimentally~\cite{C8CC01982G}.
In case of halogen substitution viz. Br and I, it helps to reduce the band gap to 2.10 and 1.85 eV, respectively (see Table S2 in SI). In the aforementioned cases, while reducing the band gap, the direct gap nature remains intact, except for Co(III) substitution. We have also calculated the band gap using G$_0$W$_0$@HSE06, which is overestimated and nearer to the experimental value. In case of pristine Cs$_2$AgInCl$_6$, the band gap is 3.77 eV. The comparison of band gap computed using PBE, HSE06, and G$_0$W$_0$@HSE06 can be seen from Table S3 in SI.

The reduction in band gap can be explained by observing the atom projected partial density of states (pDOS) (see Fig.~\ref{fig_dos}). In the pristine Cs$_2$AgInCl$_6$, the Cl p-orbitals and Ag d-orbitals contribute to the VBM, whereas the In and Ag s-orbitals contribute to the CBm (see Fig.~\ref{fig_dos}(a)). In case of substitutional alloying of Cu(I) and Au(I), their d-orbitals are at higher energy level than the d-orbitals of Ag, thereby, reducing the band gap by elevating the VBM (see Fig.~\ref{fig_dos}(c) and~\ref{fig_dos}(d)). However, in case of the M(III) substitution at In site, generally, the states lie inside the valence band (VB) or the conduction band (CB), and thus, do not reduce the band gap effectively. From Fig.~\ref{fig_dos}(b), we can see that the Ga(III) is reducing the band gap by a negligible amount (as the states contributed by the Ga are lying inside VB and CB).  Whereas, the Co(III) and Ir(III) substitution at In site show a finite decrease in the band gap. This is due to the Co d-orbitals and Ir d-orbitals contribution at CBm and VBM, respectively (see Fig. S4). In case of M(II), there is a little contribution from d- and s-orbitals of M(II) at VBM and CBm, respectively, and therefore, reducing the band gap by introducing the shallow states (see Fig.~\ref{fig_dos}(e) and~\ref{fig_dos}(f)). Moreover, the Mn states are asymmetric (with respect to spin states), which indicates that Zn(II) will be more stable than Mn(II). This can also be seen from the more negative value of $\Delta H_\textrm{D}$ for Zn(II) in comparison to Mn(II). The band gap reduction on substituting Br/I at Cl site is occurred by elevating the VBM, which is due to Br/I p-orbitals contribution at VBM (see Fig.~\ref{fig_dos}(g) and~\ref{fig_dos}(h)). The reduction in band gap on mixing the halides is in line with the previous studies~\cite{doi:10.1021/acs.jpclett.6b02682,doi:10.1021/acs.chemmater.5b04231}. In all these cases, the defect levels are shallow, which is a desirable property for optoelectronic devices. Shallow defect states ensure that the recombination of photogenerated charge carriers is not prominent and thus, the decrement in charge carrier mobility and diffusion will be insignificant.

\subsection{Optical Properties}
To obtain the optical properties, that are crucial for the perovskite to be used in optoelectronic devices, we have calculated the frequency dependent complex dielectric function $\upvarepsilon(\omega) = \textrm{Re}\, (\upvarepsilon (\omega)) + \textrm{Im}\, (\upvarepsilon (\omega))$ using G$_0$W$_0$@HSE06 [the results obtained by HSE06 xc functional are shown in Fig. S5].
%which are comparable with G$_0$W$_0$@HSE06~\cite{C9TC05002G,Pela_2015}).
\begin{figure}[h]
	\centering
	\includegraphics[width=0.45\textwidth]{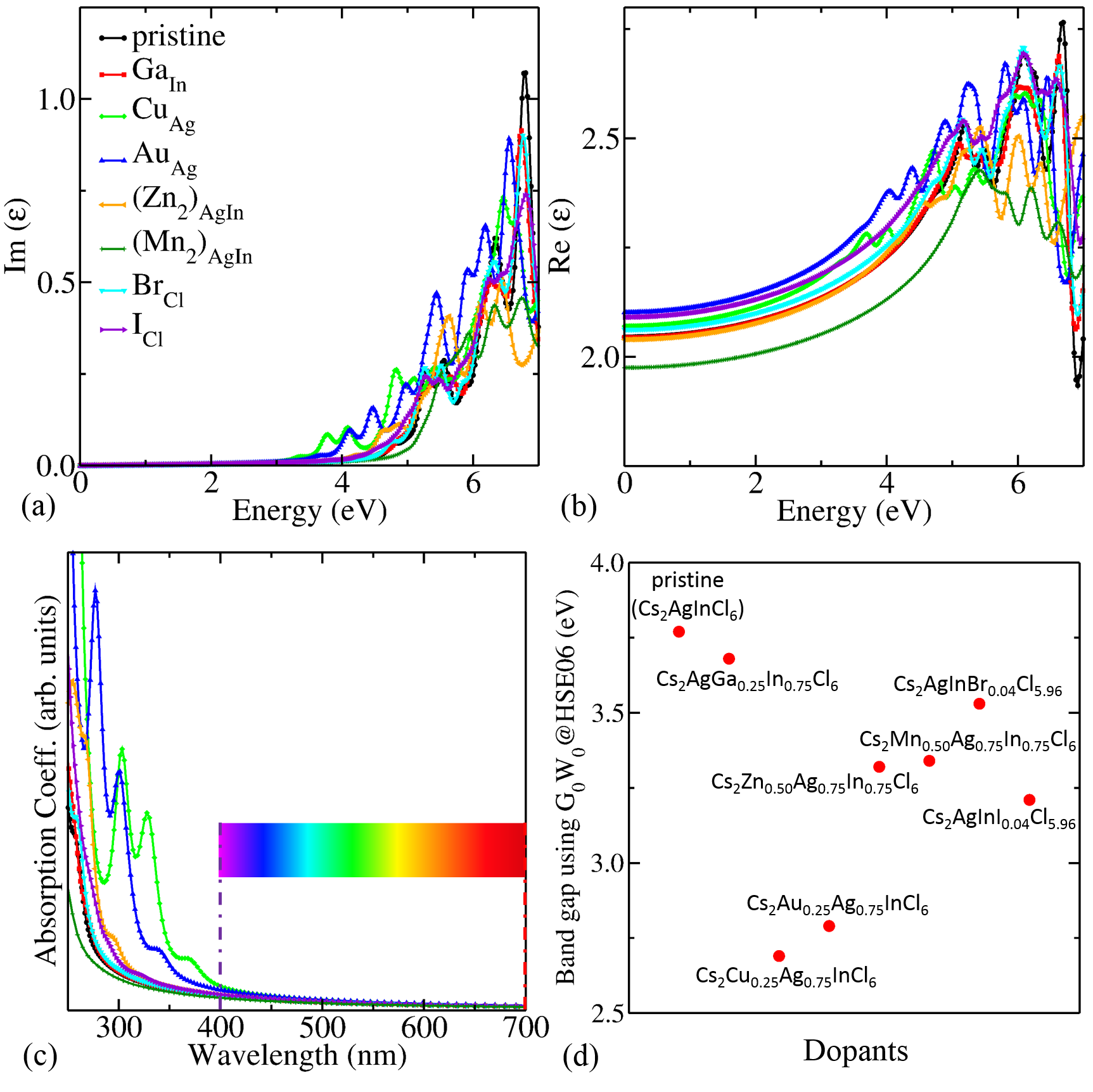}
	\caption{Spatially average (a) imaginary [Im $(\upvarepsilon)$] and (b) real [Re $(\upvarepsilon)$] part of the dielectric function, (c) absorption coefficient and (d) band gap obtained by G$_0$W$_0$@HSE06 for the pristine Cs$_2$AgInCl$_6$ and other mixed sublattices.}
	\label{fig_optical}
\end{figure}
Fig. ~\ref{fig_optical}(a) and ~\ref{fig_optical}(b) show the imaginary [Im ($\upvarepsilon$)] and real [Re ($\upvarepsilon$)] part of the dielectric function, respectively. The real static part (at $\omega=0$) of the dielectric function is a direct measure of refractive index. Higher the refractive index, better will be the probability to absorb light. On alloying, the refractive index is increased (range: $1.98-2.15$), and thus, the optical properties are enhanced. The static Re ($\upvarepsilon$) is 2.05 for pristine Cs$_2$AgInCl$_6$, and the value has increased on alloying (see Fig.~\ref{fig_optical}(b), Fig. S6 and Table S4 in SI). % This increased value of high frequency dielectric constant (Fig. 4(b)) indicates more effective screening of electron-hole interactions without phonon contribution in comparison to pristine.
The imaginary part reflects the transitions from occupied to unoccupied bands. The absorption edge is red shifted and hence, the visible light response has been achieved upon alloying (see Fig.~\ref{fig_optical}(a)). The absorption spectra have also been obtained that corroborates with the red shift observed (see Fig.~\ref{fig_optical}(c)). The absorption coefficient $\alpha(\omega)$ is related to the dielectric function as follows:
\begin{equation}
\alpha(\omega) = \sqrt{2}\,\omega\left(\sqrt{\textrm{Re} (\upvarepsilon(\omega))^2 + \textrm{Im} (\upvarepsilon(\omega))^2} - \textrm{Re} (\upvarepsilon(\omega))\right)^{\frac{1}{2}}\textrm{.}
\end{equation}
This visible response is attributed to the reduction in band gap as shown in Fig.~\ref{fig_optical}(d). The other optical parameters such as refractive index, extinction coefficient, reflectivity, optical conductivity, and energy loss spectrum have also been obtained from the dielectric tensor (see Section XII in SI).

\section{Conclusions}
In conclusion, we have investigated the role of metals M(I), M(II), M(III) and halogen X in Cs$_2$AgInCl$_6$ with mixed sublattices for inducing the visible light response by tuning its electronic properties, using state of the art DFT and beyond DFT methods. %We have found that the Goldschmidt tolerance factor and octahedral factor lie in the suitable range to form stable perovskite structure. The decomposition energy of all the mixed sublattices is negative with respect to the binary decomposition indicating that the alloyed systems will not decompose into their binary precursors and thus, are thermodynamically stable. However, Co(II), Ni(II), and Cu(II) are thermodynamically unstable and could decompose into ternary compounds. Moreover, Cu(I) becomes unstable with increment in its concentration, suggesting that the alloyed system would become unstable on increasing the concentration if the difference between sizes of substitutional cations/anion and pristine's cations/anion is large. We have observed that SOC effect is negligible in the double perovskite Cs$_2$AgInCl$_6$ as it does not contain any heavy element like Lead (Pb) and Bismuth (Bi). 
Many partially substituted configurations help to tune the band gap, thereby increasing the absorption. We have inferred that the sublattices with Cu(I) and Au(I) at Ag site, Ir(III) at In site, Zn(II) at Ag and In site simultaneously, Mn(II) at Ag and In site simultaneously, and Br and I substitution at Cl site have tuned the band gap in the visible region. Hence, these can be considered as the most promising candidates for various optoelectronic devices viz. tandem solar cells, LEDs, photodetectors and photocatalysts.\\
\section*{Supplementary Material}
See supplementary material for the details of decomposition pathways, band gap and various optical properties of pristine Cs$_2$AgInCl$_6$ and other mixed sublattices.\\
\section*{Acknowledgements}
MK acknowledges CSIR, India, for the senior research fellowship [grant no. 09/086(1292)/2017-EMR-I]. MJ acknowledges CSIR, India, for the senior research fellowship [grant no. 09/086
(1344)/2018-EMR-I]. AS acknowledges IIT Delhi for the financial support. SB acknowledges the financial support from SERB under core research grant (grant no. CRG/2019/000647). We acknowledge the High Performance Computing (HPC) facility at IIT Delhi for computational resources.\\
%\section*{DATA AVAILABILITY}
%The data that support the findings of this study are available from the corresponding author upon reasonable request.
%The data that support the findings of this study are available from the corresponding author upon reasonable request.
%\section*{References}
\nocite{*}
%\bibliography{apl}% Produces the bibliography via BibTeX.
%merlin.mbs aipnum4-1.bst 2010-07-25 4.21a (PWD, AO, DPC) hacked
%Control: key (0)
%Control: author (8) initials jnrlst
%Control: editor formatted (1) identically to author
%Control: production of article title (0) allowed
%Control: page (1) range
%Control: year (1) truncated
%Control: production of eprint (0) enabled
%
\end{document}

\preprint{AIP/123-QED}
	
\title{Sublattice mixing in Cs$_2$AgInCl$_6$ for enhanced optical properties from first-principles}
	% Force line breaks with \\
	
\author{Manish Kumar}
\email{manish.kumar@physics.iitd.ac.in}
	% \altaffiliation[Also at ]{Physics Department, XYZ University.}%Lines break automatically or can be forced with \\
\author{Manjari Jain}
\author{Arunima Singh}
\author{Saswata Bhattacharya}%
\email{saswata@physics.iitd.ac.in}
\affiliation{Department of Physics, Indian Institute of Technology Delhi, New Delhi 110016, India%\\This line break forced with \textbackslash\textbackslash
}%
	
%\author{C. Author}
% \homepage{http://www.Second.institution.edu/~Charlie.Author.}
%\affiliation{%
%Second institution and/or address%\\This line break forced% with \\
%}%
	
%\date{\today}% It is always \today, today,
%  but any date may be explicitly specified
\maketitle
%%%%%%%%%%%%%%%%%
\begin{center}
{\Large \bf Supplemental Material}\\ 
\end{center}
\begin{enumerate}[\bf I.]
	\item Band gap convergence with respect to the number of bands using G$_0$W$_0$@PBE for Cs$_2$AgInCl$_6$
	\item Radial distribution function to see octahedral distortion in Cs$_2$Cu$_{0.25}$Ag$_{0.75}$InCl$_6$
	\item Change in band gap on increasing concentration of the dopants
	\item Tolerance factor, octahedral factor, band gap and decomposition enthalpy (decomposition into binary compounds) using PBE and HSE06 xc functional of (un)doped Cs$_2$AgInCl$_6$ 
	\item Reactions for decomposition of pristine and doped Cs$_2$AgInCl$_6$ into binary/ternary compounds
	\item Spin-orbit coupling (SOC) effect in Cs$_2$Au$_{0.25}$Ag$_{0.75}$InCl$_6$
	\item Band gap computed using PBE, HSE06, and G$_0$W$_0$@HSE06 for different configurations
	\item Density of states of Cs$_2$AgCo$_{0.25}$In$_{0.75}$Cl$_6$ and Cs$_2$AgIr$_{0.25}$In$_{0.75}$Cl$_6$
	\item Dielectric function of (un)doped Cs$_2$AgInCl$_6$ using HSE06 xc functional
	\item Dielectric function of Cs$_2$AgCo$_{0.25}$In$_{0.75}$Cl$_6$ and Cs$_2$AgIr$_{0.25}$In$_{0.75}$Cl$_6$ using G$_0$W$_0$@HSE06
	\item Dielectric constant ($\upvarepsilon_\infty$) of un(doped) Cs$_2$AgInCl$_6$ using G$_0$W$_0$@HSE06
	\item Optical properties of pristine Cs$_2$AgInCl$_6$ and other mixed sublattices using G$_0$W$_0$@HSE06
\end{enumerate}
\vspace*{12pt}
\clearpage
%%%%%%%%%%%%%%
\noindent{\textbf{Band gap convergence with respect to the number of bands using G$_0$W$_0$@PBE for Cs$_2$AgInCl$_6$}\\
The pristine Cs$_2$AgInCl$_6$ has 168 occupied bands. With increasing number of unoccupied bands, the band gap gets converged. 
\begin{table}[H]
	\caption{Band gap evolution with respect to the number of bands using G$_0$W$_0$@PBE of Cs$_2$AgInCl$_6$}
	\label{tbl:x}
	\centering
	\setlength\extrarowheight{+4pt}
	\begin{tabular}{p{4.5cm}p{1.6cm}}
		\hline
		Number of bands & Band gap (eV) \\
		\hline
		240 &3.71\\
		480 &2.95\\
		640 &2.77\\
		800 &2.65\\
		\hline
	\end{tabular}
\end{table} 
\newpage
\noindent{\textbf{Radial distribution function to see octahedral distortion}\\
We have not observed any octahedral tilting, as the structures remain cubic on mixing the ions. The difference of Goldschmidt tolerance factor in comparison to pristine is only 0.03, which do not likely induce the tilting. The only thing that has minutely changed is the bond lengths B-X/B$'$-X ($\sim$0.1 \AA ). It is further supported by plotting the radial distribution function of the octahedral units of pristine and doped system. We have shown here for a test case viz. Cs$_2$Cu$_{0.25}$Ag$_{0.75}$InCl$_6$, in which the octahedral unit just stretches inwards on doping Cu (25\%) at Ag site (see Figure~\ref{fig_rad}). Similar is the case with other dopants as well, where there is change in the bond lengths of octahedra by $\sim \pm 0.1$ \AA. Hence, the distortion is negligible. \\\\
Also, it should be noted that these structures are considerable at room temperatures. This can be understood by the fact that, the synthesis of halide perovskites
is often performed near room temperature, where the 0 K enthalpy of solid-state compounds is a good approximation to predict the thermodynamic stability.  Hence, thermodynamic stability along with negligible octahedral distortion indicates the stable cubic structure at room temperature as well.\\\\
\begin{figure}[H]
	\centering
	\includegraphics[width=0.7\textwidth]{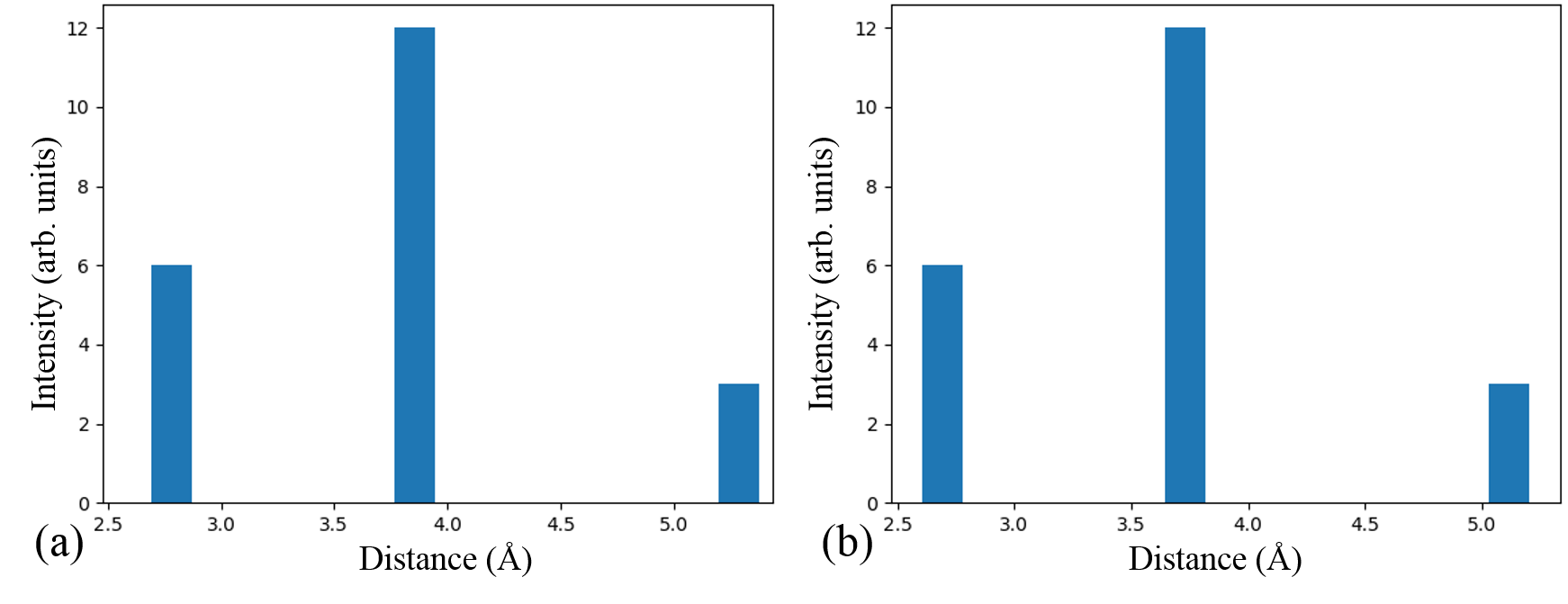}
	\caption{Radial distribution function of (a) AgCl$_6$ octahedral unit of Cs$_2$AgInCl$_6$, and (b) CuCl$_6$ octahedral unit of Cs$_2$Cu$_{0.25}$Ag$_{0.75}$InCl$_6$.}
	\label{fig_rad}
\end{figure}
\newpage
\begin{figure}[H]
	\centering
	\includegraphics[width=0.5\textwidth]{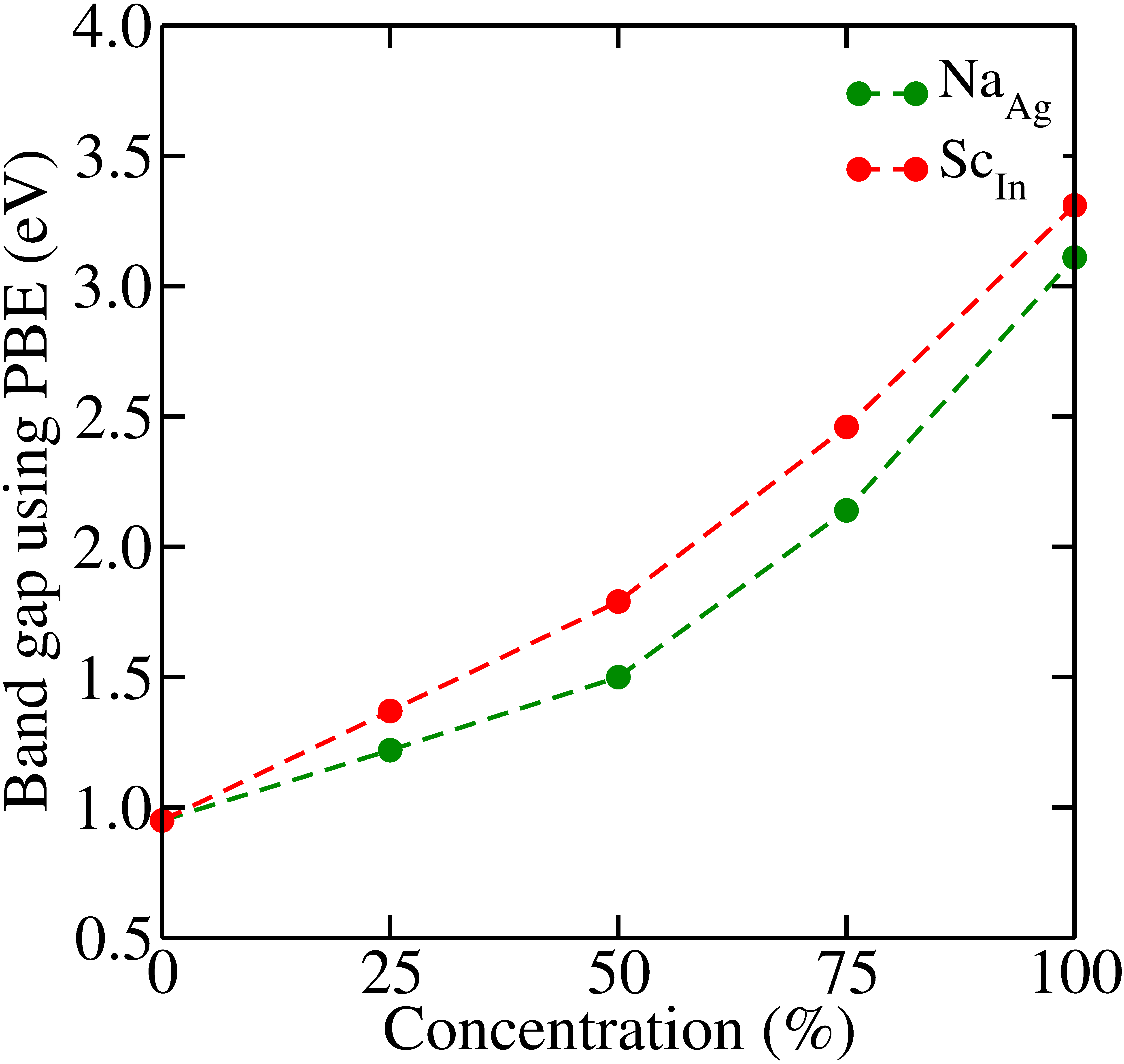} 
	\caption{Change in band gap on increasing concentration of the impurity atoms.}
	\label{fig_conc}
\end{figure}
\newpage

\begin{table}[H]
	\caption{Tolerance factor, octahedral factor, band gap and decomposition enthalpy (decomposition into binary compounds) using PBE and HSE06 xc functional of different configurations}
	\label{tbl:ex}
	\centering
	\setlength\extrarowheight{-7pt}
	\begin{tabular}{p{4.5cm} p{1.5cm}p{1.8cm}p{1.5cm}p{1.6cm}p{1.5cm}p{1.6cm}}
		\hline
		Configuration & Tolerance factor ($t$) & Octahedral factor ($\mu$) & Band gap (PBE) (eV) & $\Delta\textrm{H}_\textrm{D}$ (binary decomposition) (PBE) (eV/atom)& Band gap HSE06 (eV)& $\Delta\textrm{H}_\textrm{D}$ (binary decomposition) (HSE06) (eV/atom)\\
		\hline
		Cs$_2$AgInCl$_6$ & 0.88  & 0.54 &0.95&-0.233&2.31&-0.248\\
		Cs$_2\textrm{Cu}_{0.25}$Ag$_{0.75}\textrm{In}$Cl$_6$ & 0.90 & 0.51 &0.04&-0.224&1.51&-0.237\\
		Cs$_2\textrm{Au}_{0.25}$Ag$_{0.75}\textrm{In}$Cl$_6$  & 0.87 & 0.55&0.32&-0.212&1.54&-0.228 \\
		Cs$_2\textrm{Na}_{0.25}$Ag$_{0.75}\textrm{In}$Cl$_6$ & 0.89 & 0.53&1.22&-0.232&&\\
		Cs$_2\textrm{K}_{0.25}$Ag$_{0.75}\textrm{In}$Cl$_6$ & 0.87 & 0.55&1.37&-0.216&&\\
		Cs$_2\textrm{Rb}_{0.25}$Ag$_{0.75}\textrm{In}$Cl$_6$   & 0.87 & 0.56  &1.43&-0.200&&\\
		Cs$_2\textrm{Ag}$Al$_{0.25}\textrm{In}_{0.75}$Cl$_6$ & 0.89 & 0.52&1.03&-0.196&&\\
		Cs$_2\textrm{Ag}$Ga$_{0.25}\textrm{In}_{0.75}$Cl$_6$  & 0.89 & 0.53 &0.87&-0.192&2.23&-0.202\\
		Cs$_2\textrm{Ag}$Tl$_{0.25}\textrm{In}_{0.75}$Cl$_6$ & 0.88 & 0.54&0.37&-0.218&&\\
		Cs$_2\textrm{Ag}$Co$_{0.25}\textrm{In}_{0.75}$Cl$_6$ & 0.89 & 0.52&0.73&-0.217&2.09&-0.230\\
		Cs$_2\textrm{Ag}$Rh$_{0.25}\textrm{In}_{0.75}$Cl$_6$   & 0.89 & 0.53  &0.91&-0.195&2.53&-0.204\\
		Cs$_2\textrm{Ag}$Ir$_{0.25}\textrm{In}_{0.75}$Cl$_6$ & 0.89 & 0.53&0.43&-0.191&2.08&-0.202\\
		Cs$_2\textrm{Ag}$Sc$_{0.25}\textrm{In}_{0.75}$Cl$_6$  & 0.89  & 0.53&1.37&-0.200&&\\
		Cs$_2\textrm{Ag}$Y$_{0.25}\textrm{In}_{0.75}$Cl$_6$ & 0.88 & 0.54&1.54&-0.194&&\\
		Cs$_2\textrm{Ag}$La$_{0.25}\textrm{In}_{0.75}$Cl$_6$ & 0.87 & 0.55&1.89&-0.178&&\\
		Cs$_2\textrm{Ag}$Ac$_{0.25}\textrm{In}_{0.75}$Cl$_6$   & 0.87 & 0.56 &1.78&-0.161&&\\
		Cs$_2\textrm{Ag}$Cr$_{0.25}\textrm{In}_{0.75}$Cl$_6$ & 0.89 & 0.53&1.06&-0.197&&\\
		Cs$_2\textrm{Ag}$Sb$_{0.25}\textrm{In}_{0.75}$Cl$_6$  & 0.89 & 0.54 &1.72&-0.197&&\\
		Cs$_2\textrm{Ag}$Bi$_{0.25}\textrm{In}_{0.75}$Cl$_6$ & 0.87 & 0.55&1.96&-0.200&&\\
		Cs$_2\textrm{Ca}_{0.50}$Ag$_{0.75}\textrm{In}_{0.75}$Cl$_6$ & 0.88 & 0.54&1.89&-0.177&&\\
		Cs$_2\textrm{Sr}_{0.50}$Ag$_{0.75}\textrm{In}_{0.75}$Cl$_6$ & 0.87 & 0.57&2.11&-0.148&&\\
		Cs$_2\textrm{Ba}_{0.50}$Ag$_{0.75}\textrm{In}_{0.75}$Cl$_6$  & 0.85  & 0.59&2.29&-0.096&&\\
		Cs$_2\textrm{Zn}_{0.50}$Ag$_{0.75}\textrm{In}_{0.75}$Cl$_6$ & 0.90 & 0.51&0.61&-0.176&1.87&-0.186\\
		Cs$_2\textrm{Cd}_{0.50}$Ag$_{0.75}\textrm{In}_{0.75}$Cl$_6$ & 0.89 & 0.54&0.64&-0.180&&\\
		Cs$_2\textrm{Ge}_{0.50}$Ag$_{0.75}\textrm{In}_{0.75}$Cl$_6$   & 0.90  & 0.50 &1.43&-0.196&&\\
		Cs$_2\textrm{Sn}_{0.50}$Ag$_{0.75}\textrm{In}_{0.75}$Cl$_6$ & 0.87 & 0.57&1.18&-0.176&&\\
		Cs$_2\textrm{Pb}_{0.50}$Ag$_{0.75}\textrm{In}_{0.75}$Cl$_6$  & 0.87  & 0.57&1.72&-0.160&&\\
		Cs$_2\textrm{Mn}_{0.50}$Ag$_{0.75}\textrm{In}_{0.75}$Cl$_6$ & 0.90 & 0.52&0.31&-0.187&1.77&-0.194\\
		Cs$_2\textrm{Co}_{0.50}$Ag$_{0.75}\textrm{In}_{0.75}$Cl$_6$ & 0.90 & 0.51&metallic&-0.166&0.95&-0.140\\
		Cs$_2\textrm{Ni}_{0.50}$Ag$_{0.75}\textrm{In}_{0.75}$Cl$_6$ & 0.91 & 0.50&metallic&-0.160&1.98&-0.168\\
		Cs$_2\textrm{Cu}_{0.50}$Ag$_{0.75}\textrm{In}_{0.75}$Cl$_6$ & 0.90 & 0.50&metallic&-0.172&1.89&-0.140\\
		Cs$_2$AgInBr$_{0.04}$Cl$_{5.96}$ & 0.88  & 0.54&0.80&-0.226&2.10&-0.242\\
		Cs$_2$AgInI$_{0.04}$Cl$_{5.96}$ & 0.88  & 0.54&0.64&-0.206&1.85&-0.224\\
		\hline
	\end{tabular}
\end{table}
\newpage
\noindent{\textbf{Reactions for decomposition of pristine and doped Cs$_2$AgInCl$_6$ into binary/ternary compounds:}}\\
(1) Cs$_2$AgInCl$_6$\\
\ch{Cs8Ag4In4Cl24 -> 8 CsCl + 4 AgCl + 4 InCl3}\\
\ch{Cs8Ag4In4Cl24 -> 2 CsAgCl2 + 2 Cs3In2Cl9 + 2 AgCl}\\
(2) Cs$_2\textrm{Cu}_{0.25}$Ag$_{0.75}\textrm{In}$Cl$_6$\\
\ch{Cs8Cu1Ag3In4Cl24 -> 8 CsCl + 3 AgCl + CuCl + 4 InCl3}\\
\ch{Cs8Cu1Ag3In4Cl24 -> $\frac{1}{2}$ CsCu2Cl3 + 2 Cs3In2Cl9 + 3 AgCl + $\frac{3}{2}$ CsCl}\\
(3) Cs$_2\textrm{Au}_{0.25}$Ag$_{0.75}\textrm{In}$Cl$_6$\\
\ch{Cs8Au1Ag3In4Cl24 -> 8 CsCl + 3 AgCl + AuCl + 4 InCl3}\\
(4) Cs$_2\textrm{Na}_{0.25}$Ag$_{0.75}\textrm{In}$Cl$_6$\\
\ch{Cs8Na1Ag3In4Cl24 -> 8 CsCl + 3 AgCl + NaCl + 4 InCl3}\\
(5) Cs$_2\textrm{K}_{0.25}$Ag$_{0.75}\textrm{In}$Cl$_6$\\
\ch{Cs8K1Ag3In4Cl24 -> 8 CsCl + 3 AgCl + KCl + 4 InCl3}\\
(6) Cs$_2\textrm{Rb}_{0.25}$Ag$_{0.75}\textrm{In}$Cl$_6$\\
\ch{Cs8Rb1Ag3In4Cl24 -> 8 CsCl + 3 AgCl + RbCl + 4 InCl3}\\
(7) Cs$_2\textrm{Ag}$Al$_{0.25}\textrm{In}_{0.75}$Cl$_6$\\
\ch{Cs8Ag4Al1In3Cl24 -> 8 CsCl + 4 AgCl + 3 InCl3 + AlCl3}\\
(8) Cs$_2\textrm{Ag}$Ga$_{0.25}\textrm{In}_{0.75}$Cl$_6$\\
\ch{Cs8Ag4Ga1In3Cl24 -> 8 CsCl + 4 AgCl + 3 InCl3 + GaCl3}\\
\ch{Cs8Ag4Ga1In3Cl24 -> CsGaCl4 + $\frac{3}{2}$ Cs3In2Cl9 + 4 AgCl + $\frac{5}{2}$ CsCl}\\
(9) Cs$_2\textrm{Ag}$Tl$_{0.25}\textrm{In}_{0.75}$Cl$_6$\\
\ch{Cs8Ag4Tl1In3Cl24 -> 8 CsCl + 4 AgCl + 3 InCl3 + TlCl3}\\
(10) Cs$_2\textrm{Ag}$Co$_{0.25}\textrm{In}_{0.75}$Cl$_6$\\
\ch{Cs8Ag4Co1In3Cl24 -> 8 CsCl + 4 AgCl + 3 InCl3 + CoCl3}\\
(11) Cs$_2\textrm{Ag}$Rh$_{0.25}\textrm{In}_{0.75}$Cl$_6$\\
\ch{Cs8Ag4Rh1In3Cl24 -> 8 CsCl + 4 AgCl + 3 InCl3 + RhCl3}\\
(12) Cs$_2\textrm{Ag}$Ir$_{0.25}\textrm{In}_{0.75}$Cl$_6$\\
\ch{Cs8Ag4Ir1In3Cl24 -> 8 CsCl + 4 AgCl + 3 InCl3 + IrCl3}\\
(13) Cs$_2\textrm{Ag}$Sc$_{0.25}\textrm{In}_{0.75}$Cl$_6$\\
\ch{Cs8Ag4Sc1In3Cl24 -> 8 CsCl + 4 AgCl + 3 InCl3 + ScCl3}\\
(14) Cs$_2\textrm{Ag}$Y$_{0.25}\textrm{In}_{0.75}$Cl$_6$\\
\ch{Cs8Ag4Y1In3Cl24 -> 8 CsCl + 4 AgCl + 3 InCl3 + YCl3}\\
(15) Cs$_2\textrm{Ag}$La$_{0.25}\textrm{In}_{0.75}$Cl$_6$\\
\ch{Cs8Ag4La1In3Cl24 -> 8 CsCl + 4 AgCl + 3 InCl3 + LaCl3}\\
(16) Cs$_2\textrm{Ag}$Ac$_{0.25}\textrm{In}_{0.75}$Cl$_6$\\
\ch{Cs8Ag4Ac1In3Cl24 -> 8 CsCl + 4 AgCl + 3 InCl3 + AcCl3}\\
(17) Cs$_2\textrm{Ag}$Cr$_{0.25}\textrm{In}_{0.75}$Cl$_6$\\
\ch{Cs8Ag4Cr1In3Cl24 -> 8 CsCl + 4 AgCl + 3 InCl3 + CrCl3}\\
(18) Cs$_2\textrm{Ag}$Sb$_{0.25}\textrm{In}_{0.75}$Cl$_6$\\
\ch{Cs8Ag4Sb1In3Cl24 -> 8 CsCl + 4 AgCl + 3 InCl3 + SbCl3}\\
(19) Cs$_2\textrm{Ag}$Bi$_{0.25}\textrm{In}_{0.75}$Cl$_6$\\
\ch{Cs8Ag4Bi1In3Cl24 -> 8 CsCl + 4 AgCl + 3 InCl3 + BiCl3}\\
(20) Cs$_2\textrm{Ca}_{0.50}$Ag$_{0.75}\textrm{In}_{0.75}$Cl$_6$\\
\ch{Cs8Ca2Ag3In3Cl24 -> 8 CsCl + 3 AgCl + 3 InCl3 + 2 CaCl2}\\
(21) Cs$_2\textrm{Sr}_{0.50}$Ag$_{0.75}\textrm{In}_{0.75}$Cl$_6$\\
\ch{Cs8Sr2Ag3In3Cl24 -> 8 CsCl + 3 AgCl + 3 InCl3 + 2 SrCl2}\\
(22) Cs$_2\textrm{Ba}_{0.50}$Ag$_{0.75}\textrm{In}_{0.75}$Cl$_6$\\
\ch{Cs8Ba2Ag3In3Cl24 -> 8 CsCl + 3 AgCl + 3 InCl3 + 2 BaCl2}\\
(23) Cs$_2\textrm{Zn}_{0.50}$Ag$_{0.75}\textrm{In}_{0.75}$Cl$_6$\\
\ch{Cs8Zn2Ag3In3Cl24 -> 8 CsCl + 3 AgCl + 3 InCl3 + 2 ZnCl2}\\
\ch{Cs8Zn2Ag3In3Cl24 -> 2 Cs2ZnCl4 + $\frac{1}{2}$ Cs3In2Cl9 + 3 AgCl + $\frac{5}{2}$ CsCl + 2 InCl3}\\
(24) Cs$_2\textrm{Cd}_{0.50}$Ag$_{0.75}\textrm{In}_{0.75}$Cl$_6$\\
\ch{Cs8Cd2Ag3In3Cl24 -> 8 CsCl + 3 AgCl + 3 InCl3 + 2 CdCl2}\\
(25) Cs$_2\textrm{Ge}_{0.50}$Ag$_{0.75}\textrm{In}_{0.75}$Cl$_6$\\
\ch{Cs8Ge2Ag3In3Cl24 -> 8 CsCl + 3 AgCl + 3 InCl3 + 2 GeCl2}\\
(26) Cs$_2\textrm{Sn}_{0.50}$Ag$_{0.75}\textrm{In}_{0.75}$Cl$_6$\\
\ch{Cs8Sn2Ag3In3Cl24 -> 8 CsCl + 3 AgCl + 3 InCl3 + 2 SnCl2}\\
(27) Cs$_2\textrm{Pb}_{0.50}$Ag$_{0.75}\textrm{In}_{0.75}$Cl$_6$\\
\ch{Cs8Pb2Ag3In3Cl24 -> 8 CsCl + 3 AgCl + 3 InCl3 + 2 PbCl2}\\
(28) Cs$_2\textrm{Mn}_{0.50}$Ag$_{0.75}\textrm{In}_{0.75}$Cl$_6$\\
\ch{Cs8Mn2Ag3In3Cl24 -> 8 CsCl + 3 AgCl + 3 InCl3 + 2 MnCl2}\\
\ch{Cs8Mn2Ag3In3Cl24 -> 2 CsMnCl3 + $\frac{3}{2}$ Cs3In2Cl9 + 3 AgCl + $\frac{3}{2}$ CsCl}\\
(29) Cs$_2\textrm{Co}_{0.50}$Ag$_{0.75}\textrm{In}_{0.75}$Cl$_6$\\
\ch{Cs8Co2Ag3In3Cl24 -> 8 CsCl + 3 AgCl + 3 InCl3 + 2 CoCl2}\\
\ch{Cs8Co2Ag3In3Cl24 -> 2 CsCoCl3 + $\frac{3}{2}$ Cs3In2Cl9 + 3 AgCl + $\frac{3}{2}$ CsCl}\\
(30) Cs$_2\textrm{Ni}_{0.50}$Ag$_{0.75}\textrm{In}_{0.75}$Cl$_6$\\
\ch{Cs8Ni2Ag3In3Cl24 -> 8 CsCl + 3 AgCl + 3 InCl3 + 2 NiCl2}\\
\ch{Cs8Ni2Ag3In3Cl24 -> 2 CsNiCl3 + $\frac{3}{2}$ Cs3In2Cl9 + 3 AgCl + $\frac{3}{2}$ CsCl}\\
(31) Cs$_2\textrm{Cu}_{0.50}$Ag$_{0.75}\textrm{In}_{0.75}$Cl$_6$\\
\ch{Cs8Cu2Ag3In3Cl24 -> 8 CsCl + 3 AgCl + 3 InCl3 + 2 CuCl2}\\
\ch{Cs8Cu2Ag3In3Cl24 -> 2 CsCuCl3 + $\frac{3}{2}$ Cs3In2Cl9 + 3 AgCl + $\frac{3}{2}$ CsCl}\\
(32) Cs$_2$AgInBr$_{0.04}$Cl$_{5.96}$\\
\ch{Cs8Ag4In4Br1Cl23 -> 8 CsCl + 3 AgCl + AgBr + 4 InCl3}\\
(33) Cs$_2$AgInI$_{0.04}$Cl$_{5.96}$\\
\ch{Cs8Ag4In4I1Cl23 -> 8 CsCl + 3 AgCl + AgI + 4 InCl3}
\newpage
\noindent{\textbf{Spin-orbit coupling (SOC) effect in Cs$_2$Au$_{0.25}$Ag$_{0.75}$InCl$_6$}\\
We have checked the SOC effects for Au-doped system. On substituting 25\% Au, only the band gap is changing by 0.1 eV. The nature of the band gap still remains direct at $\Gamma$-point. There is no splitting at band edges, i.e., at conduction band minimum (CBm) or valence band maximum (VBM) (see Figure~\ref{fig_bnd}). The bands remain degenerate at the VBM and CBm. Only the VBM level is lifted by 0.1 eV. So, SOC has negligible effect in case of Au-doped systems. This corroborates with the fact that it does not contain any heavy element like Pb or Bi, where some significant effect of SOC is expected in the electronic properties of such materials.
\begin{figure}[H]
	\centering
	\includegraphics[width=0.7\textwidth]{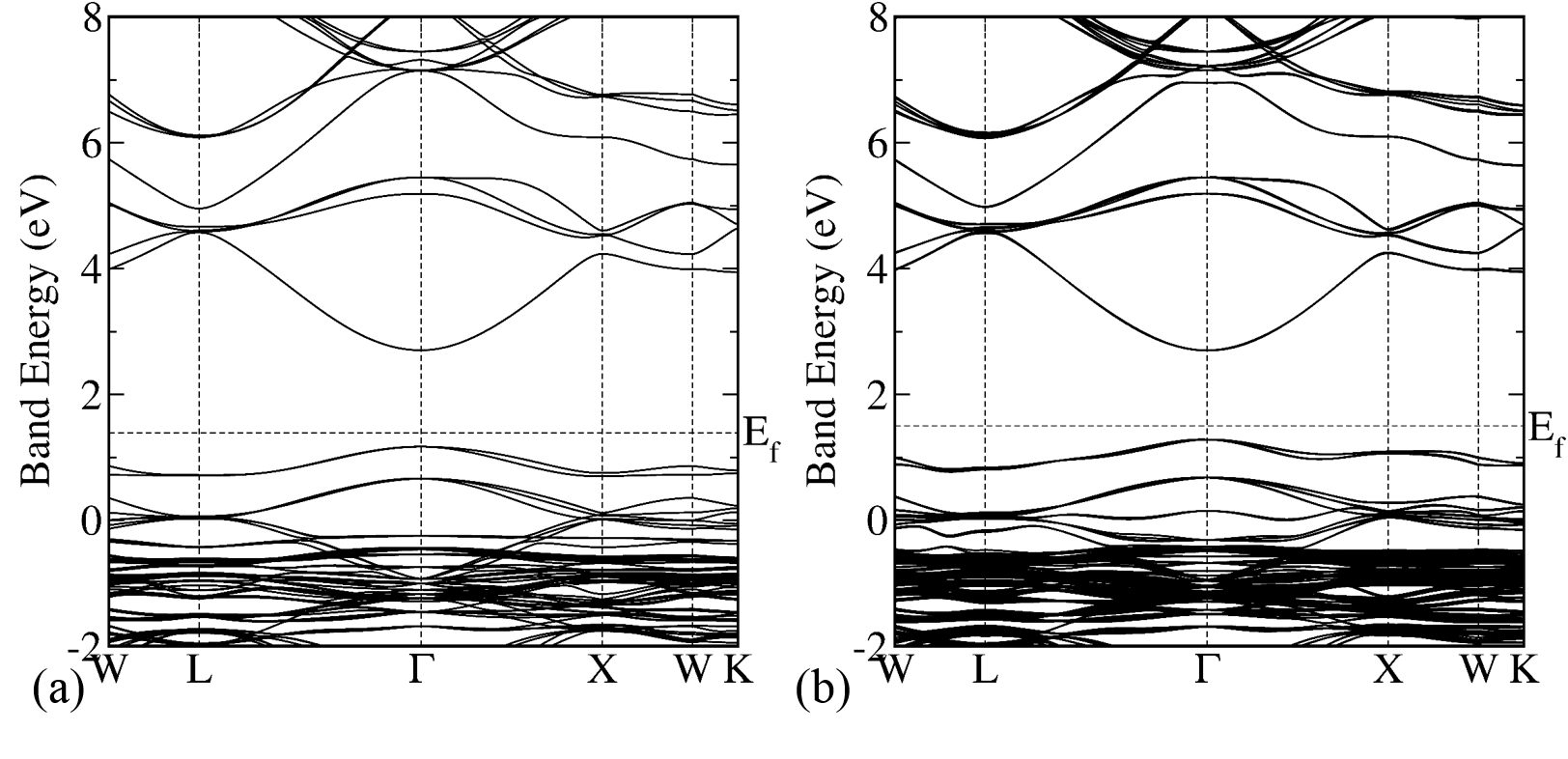}
	\caption{Bandstructure of Cs$_2$Au$_{0.25}$Ag$_{0.75}$InCl$_6$ (a) without SOC, and (b) with SOC using HSE06 functional.}
	\label{fig_bnd}
\end{figure}
\newpage
\begin{table}[H]
	\caption{Band gap (in eV) using PBE, HSE06, and G$_0$W$_0$@HSE06 for different configurations}
	\label{tbl:ex2}
	\centering
	\setlength\extrarowheight{+4pt}
	\begin{tabular}{p{4.5cm}p{1.6cm}p{1.6cm}p{2.4cm}}
		\hline
		Configuration & PBE & HSE06& G$_0$W$_0$@HSE06 \\
		\hline
		Cs$_2$AgInCl$_6$ &0.95&2.31&3.77\\
		Cs$_2\textrm{Cu}_{0.25}$Ag$_{0.75}\textrm{In}$Cl$_6$  &0.04&1.51&2.69\\
		Cs$_2\textrm{Au}_{0.25}$Ag$_{0.75}\textrm{In}$Cl$_6$ &0.32&1.54&2.79\\
		Cs$_2\textrm{Ag}$Ga$_{0.25}\textrm{In}_{0.75}$Cl$_6$   &0.87&2.23&3.68\\
		Cs$_2\textrm{Ag}$Co$_{0.25}\textrm{In}_{0.75}$Cl$_6$ &0.73&2.09&3.15\\
		Cs$_2\textrm{Ag}$Ir$_{0.25}\textrm{In}_{0.75}$Cl$_6$ &0.43&2.08&3.32\\
		Cs$_2\textrm{Zn}_{0.50}$Ag$_{0.75}\textrm{In}_{0.75}$Cl$_6$ &0.61&1.87&3.32\\
		Cs$_2\textrm{Mn}_{0.50}$Ag$_{0.75}\textrm{In}_{0.75}$Cl$_6$ &0.31&1.77&3.34\\
		Cs$_2$AgInBr$_{0.04}$Cl$_{5.96}$ &0.80&2.10&3.53\\
		Cs$_2$AgInI$_{0.04}$Cl$_{5.96}$ &0.64&1.85&3.21\\
		\hline
	\end{tabular}
\end{table}
\newpage
\begin{figure}[H]
	\centering
	\includegraphics[width=0.8\textwidth]{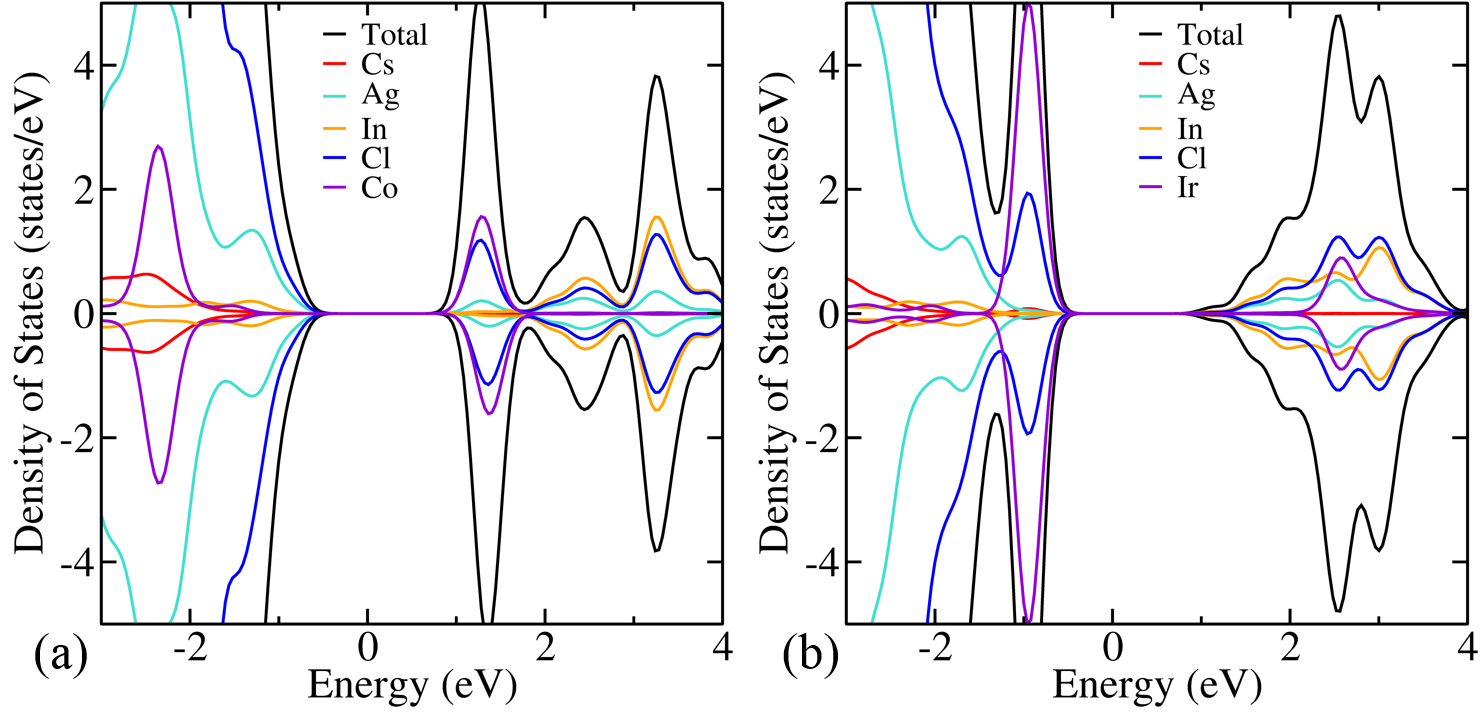} 
	\caption{Atom projected partial density of states (pDOS) using HSE06 xc functional of (a) Cs$_2$AgCo$_{0.25}$In$_{0.75}$Cl$_6$ and (b) Cs$_2$AgIr$_{0.25}$In$_{0.75}$Cl$_6$.}
	\label{fig_dos}
\end{figure}
\newpage
\begin{figure}[H]
	\centering
	\includegraphics[width=0.8\textwidth]{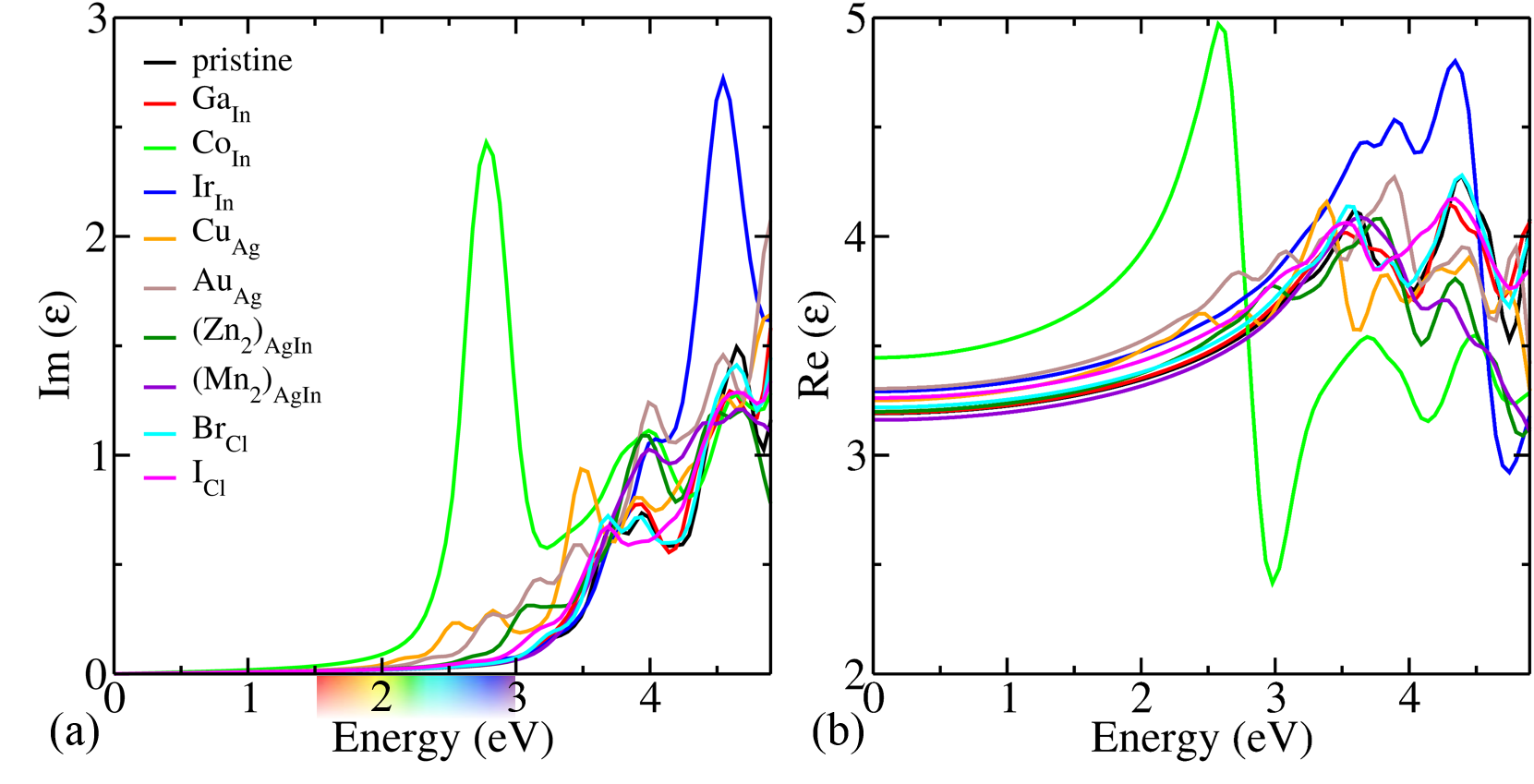}
	\caption{Spatially average (a) imaginary [Im $(\upvarepsilon)$] and (b) real [Re $(\upvarepsilon)$] part of the dielectric function obtained by HSE06 for the pristine, and doped Cs$_2$AgInCl$_6$.}
	\label{fig_hse06_opt}
\end{figure}
\newpage
\begin{figure}[H]
	\centering
	\includegraphics[width=0.8\textwidth]{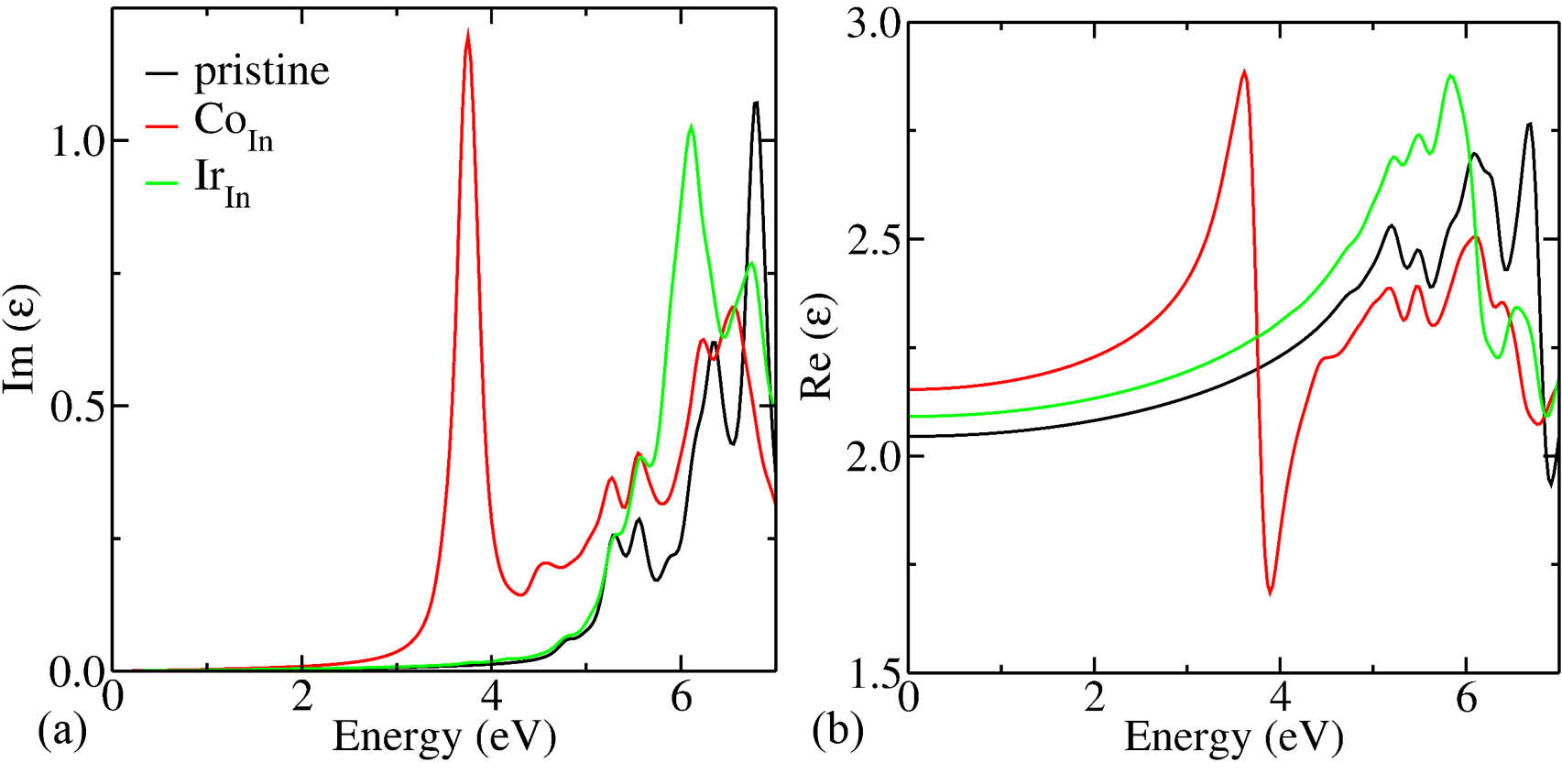} 
	\caption{Spatially average (a) imaginary [Im $(\upvarepsilon)$] and (b) real [Re $(\upvarepsilon)$] part of the dielectric function obtained by G$_0$W$_0$@HSE06 for the pristine, Cs$_2$AgCo$_{0.25}$In$_{0.75}$Cl$_6$, and Cs$_2$AgIr$_{0.25}$In$_{0.75}$Cl$_6$.}
	\label{fig_opt}
\end{figure}
\newpage
\begin{table}
	\caption{The high frequency `ion-clamped' dielectric constant ($\upvarepsilon_\infty$) using G$_0$W$_0$@HSE06}
	\label{tbl}
	\setlength\extrarowheight{-3pt}
	\begin{tabular}{ll}
		\hline
		Configurations & $\upvarepsilon_\infty$ \\
		\hline
		pristine & $2.05$\\
		Cs$_2\textrm{Ag}$Ga$_{0.25}\textrm{In}_{0.75}$Cl$_6$ & $2.04$\\
		Cs$_2\textrm{Ag}$Co$_{0.25}\textrm{In}_{0.75}$Cl$_6$ & $2.15$\\
		Cs$_2\textrm{Ag}$Ir$_{0.25}\textrm{In}_{0.75}$Cl$_6$ & $2.09$\\
		Cs$_2\textrm{Cu}_{0.25}$Ag$_{0.75}\textrm{In}$Cl$_6$ & $2.07$\\
		Cs$_2\textrm{Au}_{0.25}$Ag$_{0.75}\textrm{In}$Cl$_6$ & $2.10$\\
		Cs$_2\textrm{Zn}_{0.50}$Ag$_{0.75}\textrm{In}_{0.75}$Cl$_6$ & $2.04$\\
		Cs$_2\textrm{Mn}_{0.50}$Ag$_{0.75}\textrm{In}_{0.75}$Cl$_6$ & $1.98$\\
		Cs$_2$AgInBr$_{0.04}$Cl$_{5.96}$ & $2.06$\\
		Cs$_2$AgInI$_{0.04}$Cl$_{5.96}$ & $2.09$\\
		\hline
	\end{tabular}
\end{table}
\vspace*{12pt}
\newpage
\noindent{\textbf{Optical properties of pristine Cs$_2$AgInCl$_6$ and other mixed sublattices}}\\
The optical parameters viz. refractive index $(\eta)$, extinction coefficient $(\kappa)$, reflectivity $(R)$, absorption coefficient $(\alpha)$, optical conductivity $(\sigma)$, and energy loss spectrum $(L)$, are related to dielectric function $(\upvarepsilon)$ as follow:\\
\begin{equation}
\eta=\frac{1}{\sqrt{2}}\left[\sqrt{{\textrm{Re} (\upvarepsilon)}^2+{\textrm{Im} (\upvarepsilon)}^2}+\textrm{Re} (\upvarepsilon)\right]^{\frac{1}{2}}
\end{equation}
\begin{equation}
\kappa=\frac{1}{\sqrt{2}}\left[\sqrt{{\textrm{Re} (\upvarepsilon)}^2+{\textrm{Im} (\upvarepsilon)}^2}-\textrm{Re} (\upvarepsilon)\right]^{\frac{1}{2}}
\end{equation}
\begin{equation}
R=\frac{\left[\eta-1\right]^2+\kappa^2}{\left[\eta+1\right]^2+\kappa^2}
\end{equation}
\begin{equation}
\alpha=\frac{2\omega\kappa}{c}
\end{equation}
\begin{equation}
\sigma=\frac{\alpha\eta c}{4\pi}
\end{equation}
\begin{equation}
L=\frac{\textrm{Im} (\upvarepsilon)}{{\textrm{Re} (\upvarepsilon)}^2+{\textrm{Im} (\upvarepsilon)}^2}
\end{equation}
\newline
From Fig. \ref{fig_opt2}(a), we observe that the static value of refractive index increases on alloying, which is in direct relation with the real part of dielectric tensor. The increment in high frequency dielectric constant (real part of dielectric tensor) indicates more effective screening of electron-hole interactions without phonon contribution in comparison to pristine. This increment is due to reduction in band gap on alloying in comparison to pristine Cs$_2$AgInCl$_6$. The value of refractive index increases and has a maximum peak at first transition energy (for transition of an electron from valence band to conduction band). The maxima are shifted to lower values of energies, indicating the transition will occur at lower energies. There is abnormal behavior in case of Co(III) substitutional alloying, which is due to the indirect nature of the band gap. In this case, there will be large phonon scattering. Fig. \ref{fig_opt2}(b) shows the extinction coefficient variation with photon energy. It is a measure of absorption in materials. The smaller values of extinction coefficient indicate weaker absorption in the mixed sublattices at lower energies. Further, the values of reflectivity in these materials are smaller, that suggest their usage in transparent coatings (see Fig. \ref{fig_opt2}(c)). The peaks of the absorption coefficient and optical conductivity lie at the same values as that of the imaginary part of dielectric tensor (see Fig. \ref{fig_opt2}(d) and  \ref{fig_opt2}(e)). The absorption coefficient lies at lower energies in comparison to pristine, which implies that alloyed systems can absorb visible light of the solar spectrum. It is of the order $\sim$ 10$^{4}$ cm$^{-1}$ in lower energy region for the mixed sublattices. The optical conductivity behaves similar to the absorption coefficient. Fig. \ref{fig_opt2}(f) shows the loss spectrum function, which indicates the behavior of fast moving electrons, i.e., lose energy while passing through the material. The sharp peak in loss spectrum indicates the abrupt decrement in reflectivity spectrum for these materials.
\begin{figure}
	\centering
	\includegraphics[width=1.0\textwidth]{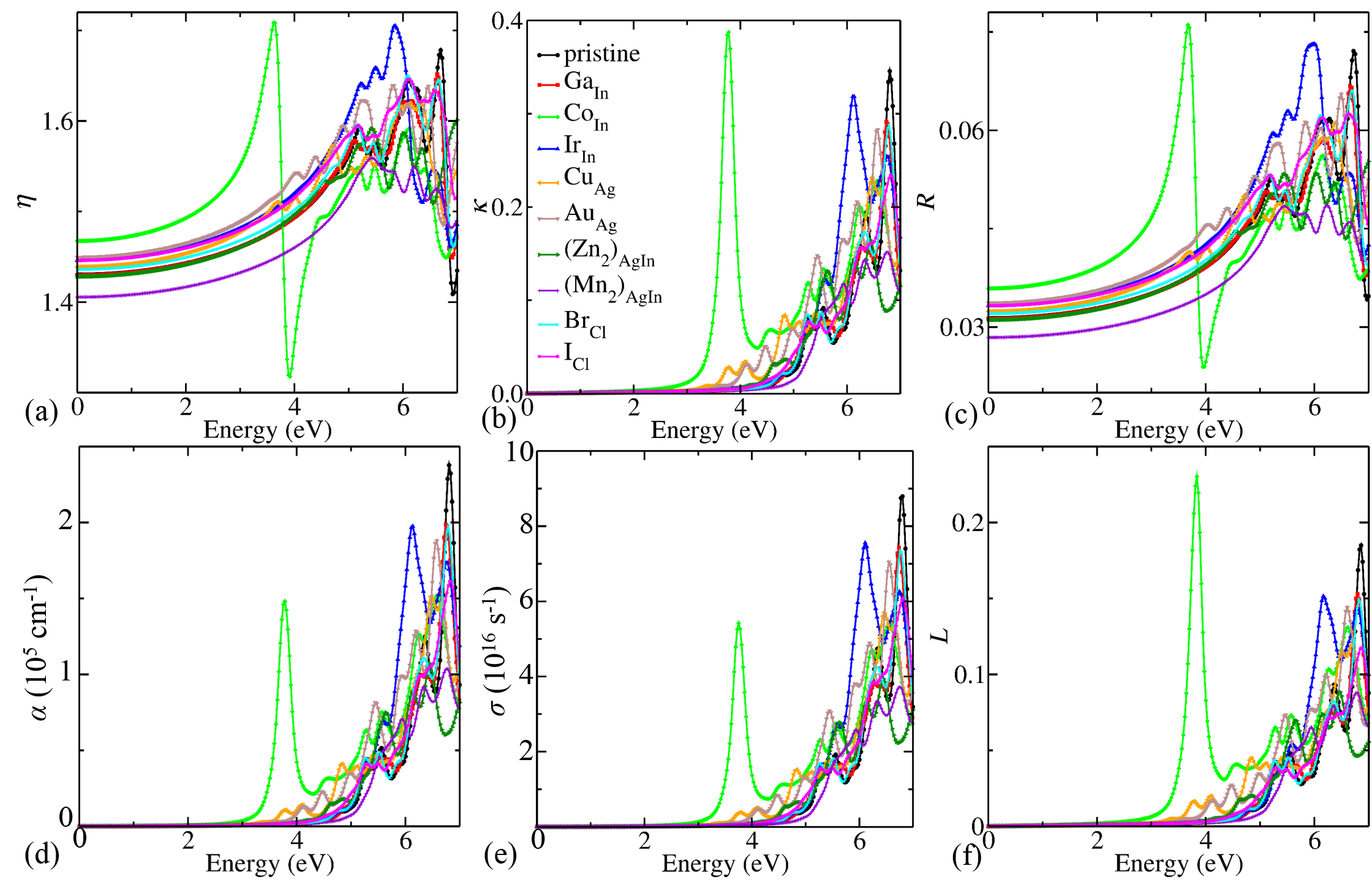} 
	\caption{Optical properties of (un)doped Cs$_2$AgInCl$_6$: (a) refractive index $(\eta)$, (b) extinction coefficient $(\kappa)$, (c) reflectivity $(R)$, (d) absorption coefficient $(\alpha)$, (e) optical conductivity $(\sigma)$, and (f) energy loss spectrum $(L)$ using  G$_0$W$_0$@HSE06.}
	\label{fig_opt2}
\end{figure}